\documentclass[aps,prb,reprint,superscriptaddress]{revtex4-1}

\usepackage{bm}
\usepackage{siunitx}
\usepackage{graphicx}
\usepackage[colorlinks=true,allcolors=blue]{hyperref}

\newcommand{\CaRuO}{Ca\textsubscript{3}Ru\textsubscript{2}O\textsubscript{7}}
\newcommand{\AFMa}{AFM\textsubscript{a}}
\newcommand{\AFMb}{AFM\textsubscript{b}}

\begin{document}

\title{Spontaneous cycloidal order mediating a spin-reorientation transition in a polar metal}

\author{C.~D.~Dashwood}
\email{cameron.dashwood.17@ucl.ac.uk}
\author{L.~S.~I.~Veiga}
\author{Q.~Faure}
\author{J.~G.~Vale}
\affiliation{London Centre for Nanotechnology and Department of Physics and Astronomy, University College London, London, WC1E 6BT, UK}

\author{D.~G.~Porter}
\author{S.~P.~Collins}
\affiliation{Diamond Light Source, Harwell Science and Innovation Campus, Didcot, Oxfordshire, OX11 0DE, UK}

\author{P.~Manuel}
\author{D.~D.~Khalyavin}
\author{F.~Orlandi}
\affiliation{ISIS Neutron and Muon Source, STFC Rutherford Appleton Laboratory, Didcot, Oxfordshire, OX11 0QX, UK}

\author{R.~S.~Perry}
\affiliation{London Centre for Nanotechnology and Department of Physics and Astronomy, University College London, London, WC1E 6BT, UK}

\author{R.~D.~Johnson}
\affiliation{Department of Physics and Astronomy, University College London, London, WC1E 6BT, UK}

\author{D.~F.~McMorrow}
\email{d.mcmorrow@ucl.ac.uk}
\affiliation{London Centre for Nanotechnology and Department of Physics and Astronomy, University College London, London, WC1E 6BT, UK}

\begin{abstract}
We show how complex modulated order can spontaneously emerge when magnetic interactions compete in a metal with polar lattice distortions. Combining neutron and resonant x-ray scattering with symmetry analysis, we reveal that the spin reorientation in \CaRuO{} is mediated by a magnetic cycloid whose eccentricity evolves smoothly but rapidly with temperature. We find the cycloid to be highly sensitive to magnetic fields, which appear to continuously generate higher harmonic modulations. Our results provide a unified picture of the rich magnetic phases of this correlated, multi-band polar metal.
\end{abstract}

\maketitle

The combination of polar distortions and magnetic order leads to the celebrated functionality of magnetoelectric multiferroics \cite{Hill2000, Spaldin2005}. Of particular interest are the magnetically-driven ferroelectrics, in which a macroscopic electric polarisation is induced by non-centrosymmetric magnetic ordering \cite{Cheong2007}. In frustrated magnets such as TbMnO\textsubscript{3} \cite{Kimura2003}, for instance, inversion symmetry is broken by complex spiral ordering which itself arises from competing magnetic interactions.

\begin{figure*}
	\includegraphics[width=\linewidth]{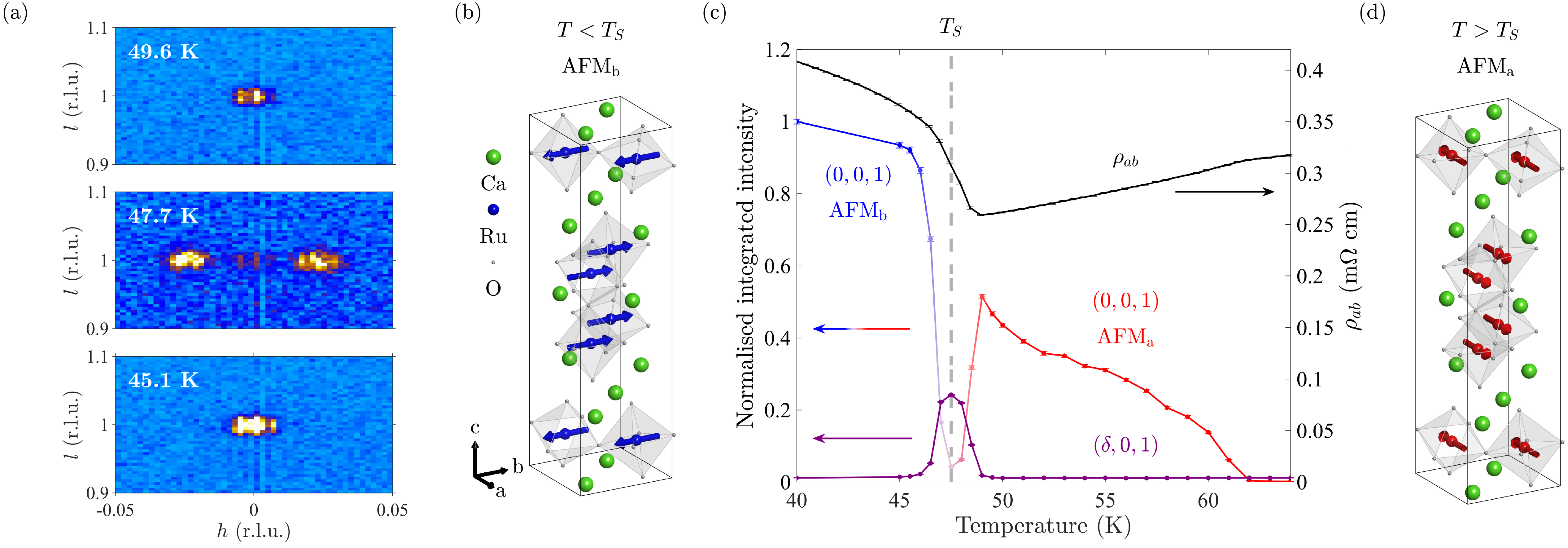}
	\caption{\label{Figure_1}Incommensurate satellites at the spin reorientation from neutron scattering. (a) Reciprocal space maps showing satellite peaks $(\pm\delta,0,1)$ around the $(0,0,1)$ magnetic peak over a narrow temperature range around $T_S =$ \SI{48}{\kelvin}. (b) Crystal and magnetic structure of \CaRuO{} below $T_S$ with collinear spins pointing along the $\bm{b}$ axis. (c) Integrated intensity of the $(0,0,1)$ (blue to red, with blue representing the \AFMb{} and red the \AFMa{} phase) and $(+\delta,0,1)$ (purple) peaks as a function of temperature, plotted alongside the in-plane resistivity $\rho_{ab}$ (black). (d) Crystal and magnetic structure of \CaRuO{} for $T_S < T < T_N$ with spins pointing along the $\bm{a}$ axis.}
\end{figure*}

While conventional multiferroics are insulating, attention has recently been focussed on polar metals, in which polar distortions and conduction electrons coexist \cite{Shi2013, Kim2016}. The bilayer ruthenate \CaRuO{} is a particularly rare example of a polar metal that also magnetically orders \cite{Cao1997, Yoshida2005, Puggioni2020}. Tilts and rotations of the RuO\textsubscript{6} unlock polar displacements of the Cu and O ions, which are not screened as only Ru contributes to the Fermi surface \cite{Lei2018}. Alongside polar domains which can be switched ferroelastically \cite{Lei2018}, \CaRuO{} demonstrates magnetic switching: a thermally driven spin-reorientation transition (SRT). At $T_S = $ \SI{48}{\kelvin} the collinear Ru moments undergo a reorientation from the $\bm{b}$ [\AFMb{}, see Fig.~\ref{Figure_1}(b)] to $\bm{a}$ axis [\AFMa{}, Fig.~\ref{Figure_1}(d)] \cite{Bohnenbuck2008, Bao2008}. In both phases the spins are aligned ferromagnetically within each bilayer and antiferromagnetically between the bilayers, with propagation vector $(0,0,1)$ \cite{Yoshida2005, Bohnenbuck2008, Bao2008}. Due to the interplay of electronic correlations and spin-orbit coupling, the SRT is intricately coupled to the crystal structure and fermiology, coinciding with an isostructural change in the lattice parameters \cite{Yoshida2005} and an increase in the resistivity caused by a gapping of most of the Fermi surface \cite{Baumberger2006, Horio2019, Markovic2020}.

In this Rapid Communication, we report the inverse of the magnetically-driven ferroelectric mechanism in \CaRuO{}: the formation of modulated order driven by competing magnetic interactions in the presence of structurally-broken inversion symmetry. Exploiting the complementarity of neutron and resonant x-ray scattering, we reveal a magnetic cycloid which evolves continuously between the collinear end states over a remarkably small temperature range to mediate the SRT. Symmetry analysis shows that the cycloid is stabilised by a uniform Dzyaloshinskii-Moriya interaction (DMI), stemming from spin-orbit coupling and activated by the small polar distortions, which competes with easy-axis anisotropies. We show how the frustration that renders the cycloid sensitive to temperature also allows it to be delicately tuned by magnetic field, under which higher harmonic modulations of the fundamental order appear to be generated. This necessitates a reinterpretation of an incommensurate structure previously observed under field \cite{Sokolov2019}, unifying it with the magnetic response to doping \cite{Ke2014, Zhu2017, Lei2019}. Our results thus demonstrate a transition at the juncture of polar metals, spin-orbit-coupled systems and magnetic frustration.

High-quality single crystals of \CaRuO{} were grown by the floating zone method, and characterised by x-ray powder diffraction, resistivity measurements and energy dispersive x-ray spectrometry. Twin domains were identified with polarised light microscopy, and single-domain pieces were cut from larger crystals using a wire saw. Measurements were performed on multiple crystals from different growth batches with consistent results. The neutron and x-ray data presented below are all from the same crystal, which was aligned by Laue diffraction. Neutron scattering measurements were performed at the WISH instrument of the ISIS Neutron and Muon Source \cite{Chapon2011}, and resonant x-ray scattering measurements at beamline I16 of the Diamond Light Source. Further experimental details can be found in the Supplemental Material \cite{supplemental}.

We first report the discovery of bulk incommensurate order in pristine \CaRuO{} from neutron scattering measurements. Figure \ref{Figure_1}(a) shows reciprocal-space maps at temperatures around $T_S$. Well above and below $T_S$ (top and bottom panels) a single peak can be seen at $\bm{q} = (0,0,1)$ reciprocal lattice units (r.l.u.) which arises from the known \AFMa{} and \AFMb{} structures. As unpolarised neutron scattering is sensitive to the component of the moment perpendicular to $\bm{q}$, we are sensitive to the full moment in both phases and the lower intensity in the former is indicative of a smaller moment, consistent with previous reports \cite{Bao2008}. Strikingly, close to $T_S$ (middle panel) the central peak is strongly suppressed, and satellites can be seen at incommensurate positions $(\pm\delta,0,1)$ with $\delta \approx 0.023$ r.l.u. Satellite peaks closely spaced around a commensurate position are indicative of a long-range modulation of the nuclear and/or magnetic structure, such as charge/spin density wave phases \cite{Tranquada1995, Lester2015} or spiral magnetic structures \cite{Ishikawa1976, Sosnowska1995}. In this case, the transfer of intensity from the commensurate to satellite peaks suggests a common magnetic origin, while their comparable widths indicate long correlation lengths. Comparing the integrated intensity of the $(0,0,1)$ and $(\delta,0,1)$ peaks to the in-plane resistivity $\rho_{ab}$ in Fig.~\ref{Figure_1}(c), it can be seen that the boundaries of the incommensurate phase correspond to changes of slope in $\rho_{ab}$, revealing how the magnetic structure is intimately linked to the electronic behaviour.

\begin{figure*}
	\includegraphics[width=\linewidth]{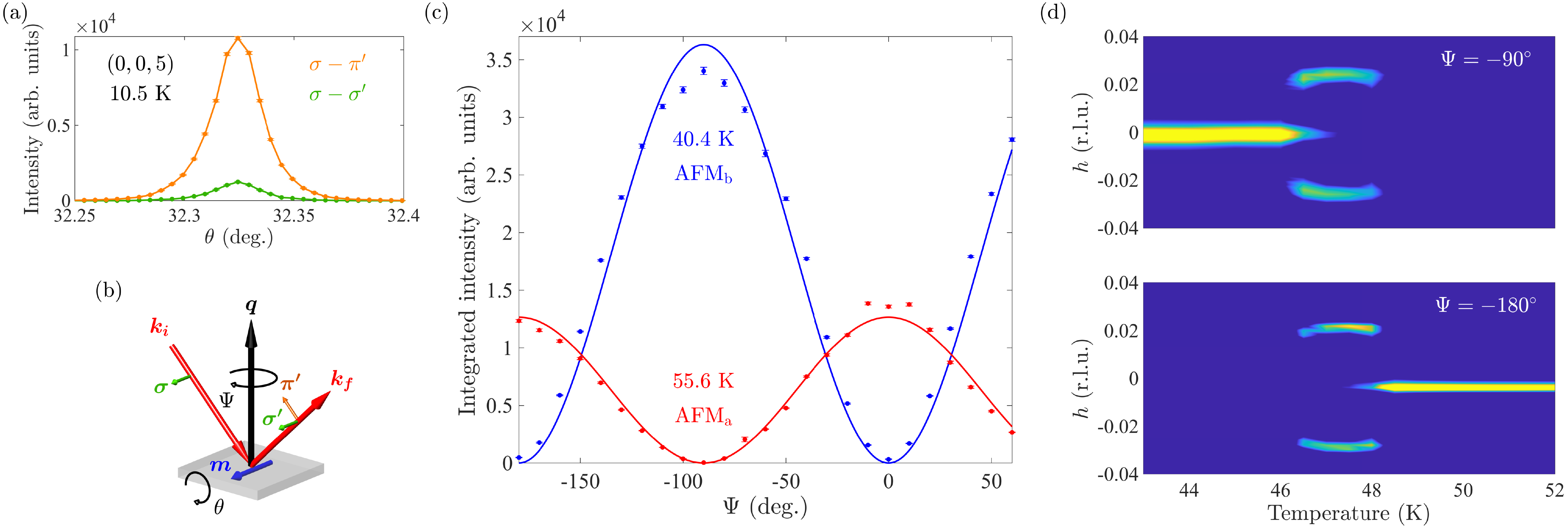}
	\caption{\label{Figure_2}Spin reorientation from resonant x-ray scattering. (a) Rocking curves of the $(0,0,5)$ peak at \SI{10.5}{\kelvin} in the $\sigma-\pi'$ (orange) and $\sigma-\sigma'$ (green) polarisation channels, confirming its magnetic origin. (b) Experimental geometry, showing incident x-rays with wavevector $\bm{k_i}$ (red arrow) scattering from a magnetic moment $\bm{m}$ (blue arrow) to wavevector $\bm{k_f}$. The incident x-rays are polarised horizontally, normal to the scattering plane ($\sigma$ polarised, green arrow), and the scattered x-rays are polarised either in the scattering plane ($\pi'$, orange) or normal it ($\sigma'$, green). The azimuth $\Psi$ is varied by rotating the sample around the scattering vector $\bm{q} = \bm{k_f} -\bm{k_i}$ (black arrow). (c) Azimuthal scans of the $(0,0,5)$ peak above (\SI{55.6}{\kelvin}, red) and below (\SI{40.4}{\kelvin}, blue) $T_S$. (d) $h$ scans through $(0,0,5)$ as a function of temperature at $\Psi =$ \ang{-90} (top panel, sensitive to the component of the moment along $\bm{b}$) and $\Psi =$ \ang{-180} (lower panel, sensitive to the component along $\bm{a}$). The intensity is plotted on a log scale.}
\end{figure*}

Having established the existence of an incommensurate phase in the vicinity of the SRT, we now turn to resonant x-ray scattering to unravel its nature. Tuning the incident x-rays to the Ru $L_2$ absorption edge causes a resonant enhancement of the scattered intensity, which combined with polarisation analysis of the scattered beam results in an element-selective probe of long-range magnetic order. Figure \ref{Figure_2}(a) shows rocking scans of the $(0,0,5)$ peak in two orthogonal polarisation channels. X-rays polarised in the scattering plane are denoted $\pi$, and those normal to it $\sigma$ [see Fig.~\ref{Figure_2}(b)]. For incident $\sigma$ x-rays, resonant scattering from a magnetic moment occurs only in the crossed $\sigma-\pi'$ polarisation channel \cite{Hill1996}, so the dominant intensity in this channel verifies the magnetic origin of the peak. The remnant intensity in the $\sigma-\sigma'$ channel is due to leakage through the analyser (see the Supplemental Material \cite{supplemental}).

In contrast to neutron scattering, resonant x-ray scattering in $\sigma-\pi'$ is sensitive to the component of the moment parallel to $\bm{k_f}$. More information about the magnetic structure can therefore be obtained by rotating the sample through an azimuthal angle $\Psi$ in order to vary the component projected along $\bm{k_f}$ [see Fig.~\ref{Figure_2}(b)]. Azimuthal dependences of the $(0,0,5)$ peak are shown in Fig.~\ref{Figure_2}(c). For basal-plane collinear structures, calculation of the resonant cross-section gives an intensity $\propto \cos^2(\Psi + \phi)$ [solid lines in Fig.~\ref{Figure_2}(c)] where $\phi$ is the rotation of the moments away from the $\bm{a}$ axis \cite{Bohnenbuck2008}. At \SI{40.4}{\kelvin} it can be seen that $\phi =$ \ang{90} and the moments are along $\bm{b}$, while at \SI{55.6}{\kelvin} $\phi =$ \ang{0} and the moments are along $\bm{a}$. The upper and lower panels in Fig.~\ref{Figure_2}(d) show $h$ scans through $(0,0,5)$ as a function of temperature for two azimuths, sensitive to the $\bm{b}$ ($\Psi =$ \ang{-90}) and $\bm{a}$ ($\Psi =$ \ang{-180}) components of the moments respectively. The spin reorientation is clearly identified by the transfer of commensurate intensity between the azimuths. The satellite peaks are also visible at both azimuths, and intriguingly have subtle temperature dependences to their wavevectors and intensities. We now analyse these satellites in more detail.

\begin{figure*}
	\includegraphics[width=\linewidth]{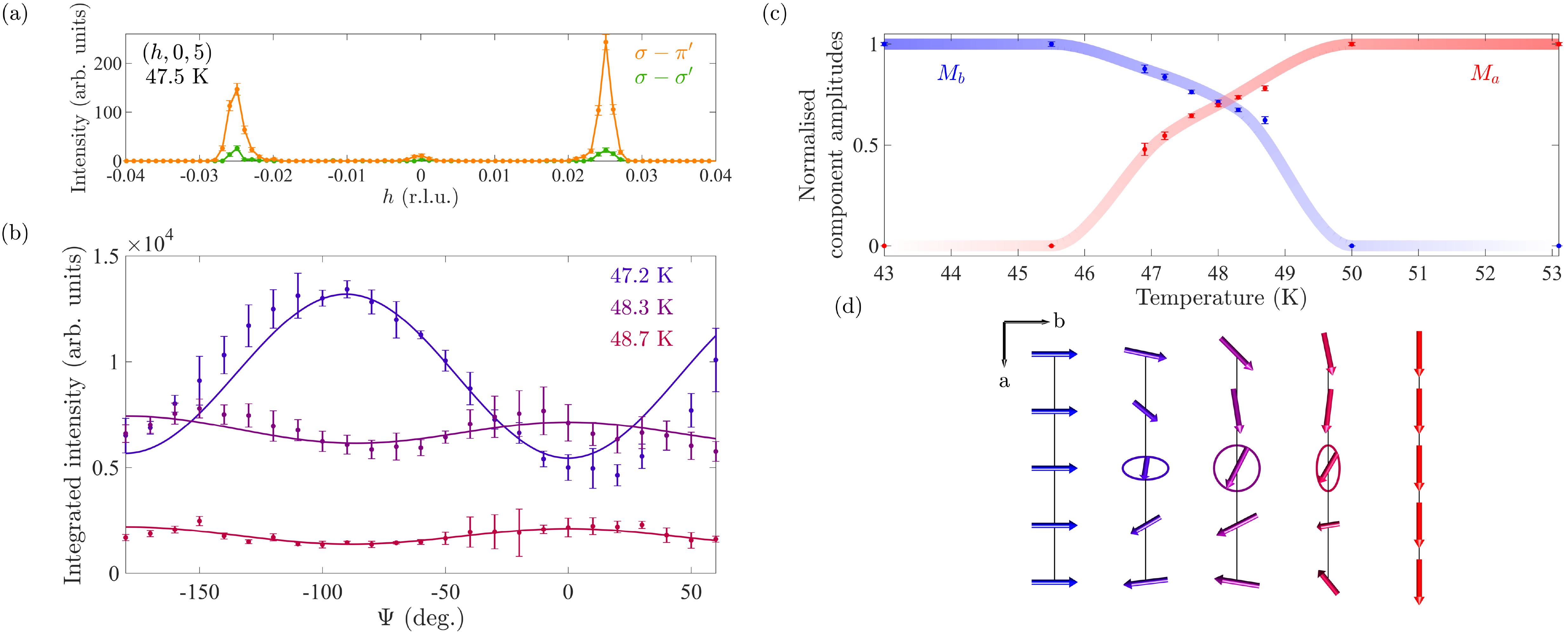}
	\caption{\label{Figure_3}Evolving cycloidal structure from resonant x-ray scattering. (a) $h$ scans through $(0,0,5)$ in both polarisation channels, showing that the incommensurate satellites are magnetic. (b) Azimuthal dependences of the $(-\delta,0,5)$ satellite at select temperatures. The solid lines are fits to the cycloidal model described in the text. (c) Amplitudes of the cycloid components extracted from fits like those in (b), normalised by $(M_a^2 + M_b^2)^{-1/2}$ to remove the effect of the increasing moment size on cooling. Solid lines are guides to the eye. (d) Schematic of the evolution of the magnetic structure with temperature, with the lengths of the arrows indicating the changing of the cycloidal envelope from elongated along $\bm{b}$, to circular, to elongated along $\bm{a}$. The changing moment size and period of the cycloid are neglected for clarity.}
\end{figure*}

Figure \ref{Figure_3}(a) shows $h$ scans in both polarisation channels, confirming the magnetic origin of the satellites. In order to determine the structure of the incommensurate phase we performed a detailed investigation of the azimuthal dependence of the satellites. Representative dependences are shown in Fig.~\ref{Figure_3}(b). All of the dependences are sinusoidal, but show dramatic changes in peak-to-peak amplitude and phase, evidencing a remarkable evolution of the structure over a small temperature window. Unlike in the commensurate phases where the intensity goes to zero when $\bm{k_f}$ is perpendicular to $\bm{m}$, here we see a finite intensity at all $\Psi$. This is a clear indication of a non-collinear rotating structure, with a component of the moment always parallel to $\bm{k_f}$. We calculated the cross-section for all possible modulated states using the \textsc{MagnetiX} package \footnote{L.~Chapon, MagneticX, \url{https://forge.epn-campus.eu/projects/magnetix}} and found that a cycloid with moments rotating in the $\bm{a}-\bm{b}$ plane, maintaining the ferromagnetic coupling within bilayers and antiferromagnetic coupling between bilayers, is uniquely consistent with our data.

To understand the temperature evolution of the magnetic structure, we developed a model in which the cycloid is decomposed into two spin-density wave components $\pi/2$ out-of-phase, that in the commensurate limits are equivalent to the \AFMb{} and \AFMa{} structures. The only free parameters are the amplitudes of these components, $M_b$ and $M_a$, which describe the elongation of the envelope of the cycloid along the $\bm{b}$ and $\bm{a}$ axes respectively. We fit the azimuthal dependences using \textsc{MagnetiX} [solid lines in Fig.~\ref{Figure_3}(b)] and found that this simple model provides a remarkably accurate description of the data at all temperatures. The fits can be intuitively understood by neglecting the small $h$ component of the wavevector, which simplifies the dependence to $\propto (M_b \sin{\Psi})^2 + (M_a \cos{\Psi})^2$. It can then be seen that the peak-to-peak amplitude of the oscillations, $M_b^2 - M_a^2$, is directly related to the eccentricity of the cycloid (after normalising out the effect of the increase in the overall moment size on cooling), while the phase depends on whether $M_b$ or $M_a$ is larger. The normalised fitted values of the amplitudes are shown in Fig.~\ref{Figure_3}(c) [including fits to the commensurate dependences such as those in Fig.~\ref{Figure_2}(c)]. In the commensurate phases only one of the components is present, as expected. In the incommensurate phase, by contrast, both amplitudes are finite and vary with temperature. This describes the magnetic structure shown schematically in Fig.~\ref{Figure_3}(d), where the envelope of the cycloid transitions from elongated along $\bm{b}$, to circular, to elongated along $\bm{a}$. Our x-ray data has therefore revealed a complex and evolving cycloidal magnetic structure which mediates the SRT.

Theoretical justification of this model is provided by a symmetry analysis of terms in the free energy (full details can be found in the Supplemental Material \cite{supplemental}). The \AFMb{} and \AFMa{} phases can each be associated with a one-dimensional order parameter, $\mu$ and $\rho$. The polar structure of \CaRuO{} allows the Lifshitz-type invariant $\mu (\partial \rho / \partial y) - \rho (\partial \mu / \partial y)$ in the free energy. Such invariants describe instabilities towards incommensurate modulated states \cite{Izyumov1990, Sosnowska1995, Kadomtseva2004}, including the cycloid observed here. The microscopic origin of the Lifshitz invariant lies in a uniform DMI, which competes with easy-axis anisotropies to select the ground state of the system. Away from $T_S$, the easy-axis anisotropies dominate and preclude the formation of a modulated state, leading to the \AFMb{} or \AFMa{} phase. As the easy-axis gradually changes from $\bm{b}$ to $\bm{a}$ in the vicinity of the SRT, however, we expect minimal or easy-plane anisotropy, allowing the uniform DMI to stabilise the cycloidal phase.

\begin{figure}
	\includegraphics[width=\linewidth]{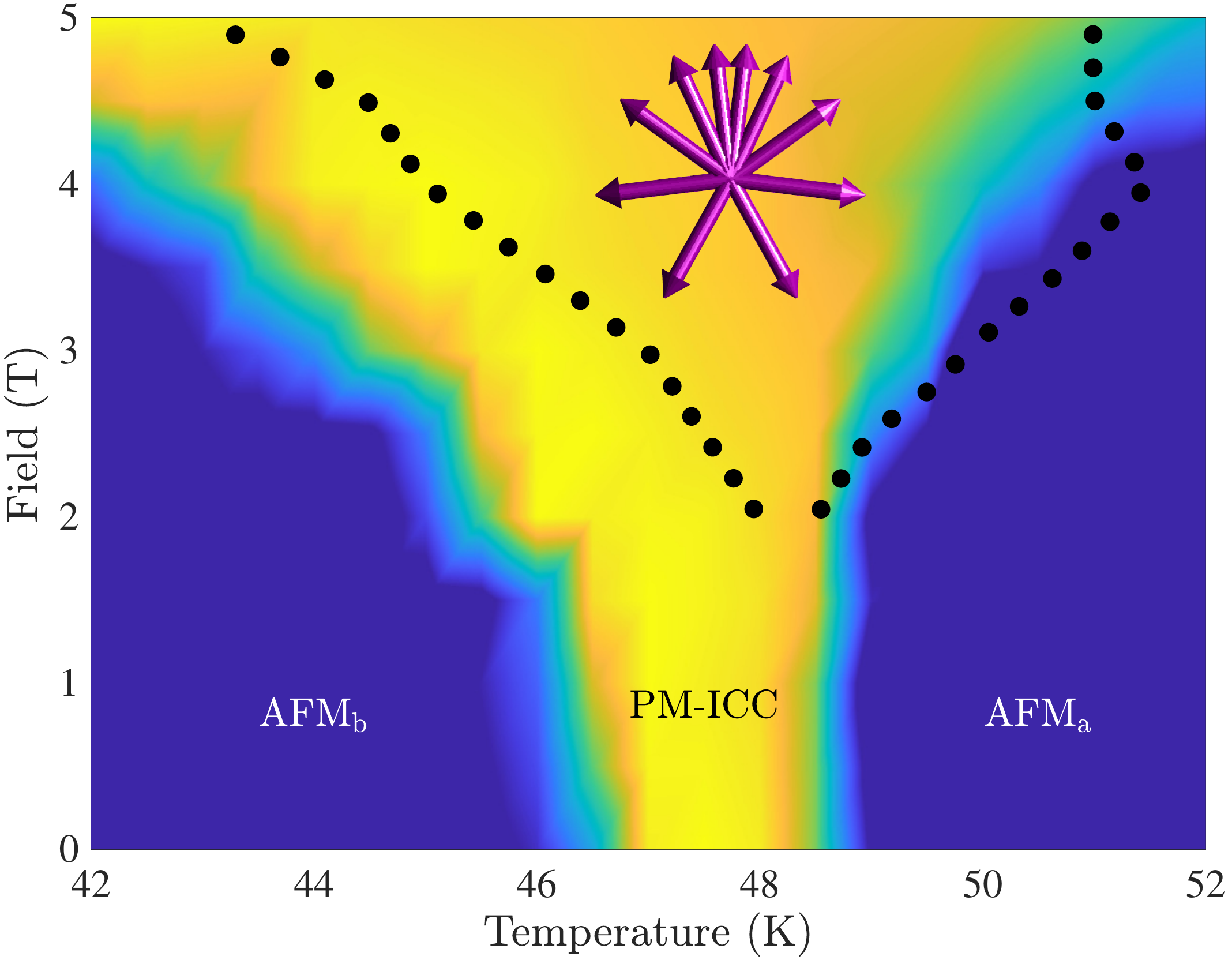}
	\caption{\label{Figure_4}Phase diagram for a magnetic field along the $\bm{b}$ axis from neutron scattering. The colour scale is the integrated intensity of the $(\delta,0,1)$ peak, with the \AFMb{}, \AFMa{} and phase-modulated incommensurate cycloid (PM-ICC) phases marked. The black dotted line encloses the region over which peaks at $(\Delta,0,0) \approx (2\delta,0,0)$ were seen in a previous SANS measurement \cite{Sokolov2019}. The purple arrows are a cartoon of the spin distribution in the $\bm{a}-\bm{b}$ plane in the PM-ICC phase, depicted with a circular envelope for clarity.}
\end{figure}

The frustration that causes the evolution of the envelope and period of the cycloid with temperature should also make it highly sensitive to other perturbations. To demonstrate this, we performed a neutron scattering experiment on \CaRuO{} with a magnetic field applied along the $\bm{b}$ axis. The resulting temperature-field phase diagram is shown in Fig.~\ref{Figure_4}, revealing a large expansion of the cycloidal phase under small fields. A striking feature of the phase boundary is its similarity with that of the so-called ``metamagnetic texture'' reported by Sokolov \textit{et al.} \cite{Sokolov2019}, who measured incommensurate peaks at $(\pm\Delta,0,0)$ in fields above \SI{2}{\tesla} with small angle neutron scattering (SANS). The connection to the magnetic satellites that we observe can be seen by doubling our wavevector $(\delta,0,1)$ and then projecting back into the first Brillouin zone through subtraction of the lattice vector $(0,0,2)$, giving $(2\delta,0,0) \approx (\Delta,0,0)$ \footnote{The wavevectors of the satellite peaks are field and temperature dependent, but comparing the value from our data at \SI{2}{\tesla} and \SI{47}{\kelvin}, \unexpanded{$\delta \approx 0.023$}, with the SANS data at at the same field and temperature, \unexpanded{$\Delta \approx 0.045$}, we can see the correspondence \unexpanded{$2\delta \approx \Delta$}}. It is then readily apparent that the peaks observed by Sokolov \textit{et al.} are in fact second harmonics of our satellites, whose presence at zero field rule out the proposed metamagnetic texture. While we could not directly observe the second harmonic satellites in our experiment \footnote{Observation of the second harmonics of the satellite peaks in our neutron experiment was precluded by the low flux of long-wavelength neutrons at such high $d$-spacing}, their appearance under magnetic fields is naturally explained by a phase-modulation of the cycloid stabilised by symmetry-allowed terms in the free energy (see the Supplemental Material \cite{supplemental}). This phase modulation corresponds to the spins bunching along the field direction in order to reduce their Zeeman energy, shown schematically by the purple arrows in Fig.~\ref{Figure_4}. In this scenario, a net magnetisation develops as higher harmonics are generated continuously from the zero-field cycloid, in the absence of any metamagnetic transition. Such behaviour is reminiscent of the highly robust, tuneable soliton lattices seen in chiral helimagnets under field \cite{Togawa2012}.

Within this framework, we can also explain previous observations of incommensurate structures in doped \CaRuO{} \cite{Ke2014, Zhu2017, Lei2019}. Here, a reduced anisotropy from the introduction of magnetic dopants should allow an easier turning of the moments away from their easy-axis by the DMI, stabilising cycloidal structures with shorter repeat distances and over larger temperature ranges.

Our analysis therefore unifies previously disparate magnetic behaviours of \CaRuO{}, attributing them to a common origin and revealing a highly rich phase diagram. Knowledge of the mediating cycloid, and the ingredients necessary for its formation from symmetry analysis, will be vital for the development of a conclusive microscopic theory of the technologically-relevant SRT. Alongside a review of bulk thermodynamic and magnetisation data in the vicinity of the SRT, recent work which ascribes the reorientation to a ``magnetoelectric'' anisotropy \cite{Markovic2020} will need revising in light of our results. Finally, given the fragility of the cycloidal order and the strong coupling between the structural, electronic and magnetic degrees of freedom in \CaRuO{}, one might imagine the possibility of using mechanical or electrical stimuli as tuning parameters, with profound applications in magnetic memory and spintronics.

\begin{acknowledgments}
We thank Frank Kr\"uger, Andrew Green, Michal Kwasigroch and Adam Walker for insightful discussions. We also thank Richard Thorogate for assistance with the resistivity measurement, Daniel Nye and Gavin Stenning for assistance with the powder x-ray diffraction and Laue alignment in the Materials Characterisation Laboratory at the ISIS Neutron and Muon Source, Mike Matthews for technical support at I16, and Jacob Simms, Katherine Mordecai, Jon Bones and David Keymer for technical support at WISH. C.~D.~D.~was supported by the Engineering and Physical Sciences Research Council (EPSRC) Centre for Doctoral Training in the Advanced Characterisation of Materials under Grant No.~EP/L015277/1. R.~D.~J.~acknowledges support from a Royal Society University Research Fellowship. Work at UCL was supported by the EPSRC under Grants No.~EP/N027671/1, EP/N034694/1 and EP/P013449/1. Experiments at the ISIS Neutron and Muon Source were supported by a beamtime allocation from the Science and Technology Facilities Council under proposal RB1920210. We acknowledge the Diamond Light Source for time on beamline I16 under proposal MM23580.
\end{acknowledgments}

\bibliography{Ca327_PRB}

\end{document}


\title{Supplemental Material for \\ Spontaneous cycloidal order mediating a spin-reorientation transition in a polar metal}

\author{C.~D.~Dashwood}
\email{cameron.dashwood.17@ucl.ac.uk}
\author{L.~S.~I.~Veiga}
\author{Q.~Faure}
\author{J.~G.~Vale}
\affiliation{London Centre for Nanotechnology and Department of Physics and Astronomy, University College London, London, WC1E 6BT, UK}

\author{D.~G.~Porter}
\author{S.~P.~Collins}
\affiliation{Diamond Light Source, Harwell Science and Innovation Campus, Didcot, Oxfordshire, OX11 0DE, UK}

\author{P.~Manuel}
\author{D.~D.~Khalyavin}
\author{F.~Orlandi}
\affiliation{ISIS Neutron and Muon Source, STFC Rutherford Appleton Laboratory, Didcot, Oxfordshire, OX11 0QX, UK}

\author{R.~S.~Perry}
\affiliation{London Centre for Nanotechnology and Department of Physics and Astronomy, University College London, London, WC1E 6BT, UK}

\author{R.~D.~Johnson}
\affiliation{Department of Physics and Astronomy, University College London, London, WC1E 6BT, UK}

\author{D.~F.~McMorrow}
\email{d.mcmorrow@ucl.ac.uk}
\affiliation{London Centre for Nanotechnology and Department of Physics and Astronomy, University College London, London, WC1E 6BT, UK}

\maketitle

\section{Crystal growth and bulk characterisation}

\begin{figure*}
	\includegraphics[width=\linewidth]{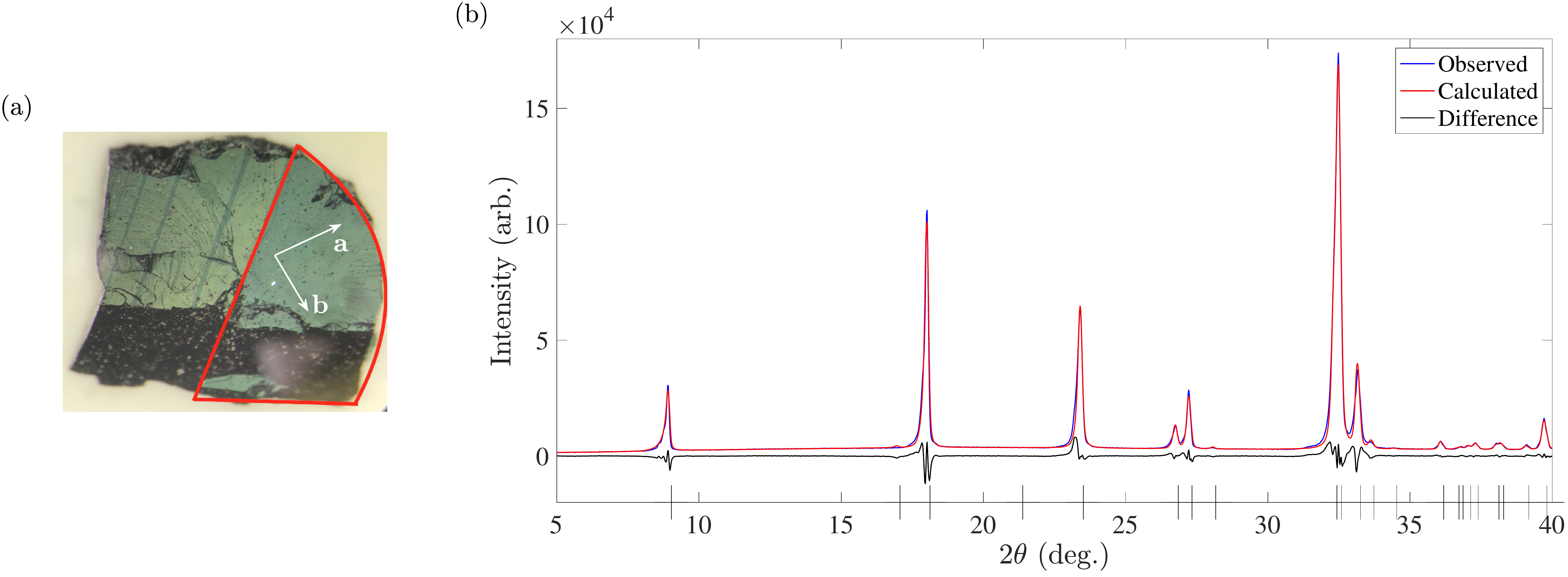}
	\caption{\label{Figure_S1}Sample characterisation. (a) Polarised light microscope image of the \CaRuO{} single crystal used in this study. Twin domains, corresponding to swapped $\bm{a}$ and $\bm{b}$ axes, are visible as light and dark stripes running diagonally along $(1,\bar{1},0)$ down the sample. The red outline marks the single-domain region that was cut out with a wire saw and used in the neutron and x-ray experiments. The crystallographic axes are marked. (b) Powder x-ray diffraction pattern of a ground crystal from the same batch as that in a, showing single-phase \CaRuO{} with no impurities. Rietveld refinement was performed in the $Bb2_1m$ space-group setting using the \textsc{Profex} software \cite{Dobelin2015} and yielded lattice parameters $a =$ \SI{5.3836(3)}{\angstrom}, $b =$ \SI{5.5174(4)}{\angstrom}, $c =$ \SI{19.5629(8)}{\angstrom}, with $R_\textrm{wp} =$ 4.12\% and $R_\textrm{exp} =$ 1.74\%. Bragg peak positions are indicated by long ticks on the horizontal axis.}
\end{figure*}

High-quality single crystals of \CaRuO{} were grown using the floating zone method in a Crystal System Corporation FZ-T-10000-H-VI-VPO-I-HR-PC mirror furnace. The crystal structure was checked using a Rigaku Miniflex 600 x-ray powder diffractometer, with the resulting diffraction pattern shown in Fig.~\ref{Figure_S1}b. The resistivity shown in Fig.~1c of the main text was measured in a standard four-probe configuration using a Quantum Design PPMS. The stoichiometry was confirmed with an Oxford Instruments energy dispersive x-ray spectrometer on a JEOL JEM-2100 electron microscope, yielding a Ca:Ru ratio of 3:1.97$\pm$0.06. A polarised light microscope was used to identify twin domains, and single-domain pieces were cut from larger crystals using a wire saw (see Fig.~\ref{Figure_S1}a).

\section{Neutron scattering}

Neutron scattering measurements were performed at the WISH instrument of the ISIS Neutron and Muon Source \cite{Chapon2011}. WISH is a time-of-flight cold neutron instrument, allowing measurement of closely spaced magnetic peaks with high $d$-spacing. The sample was mounted on the end of an aluminium pin with aluminium tape. For the detector images shown in Fig.~1a of the main text, the sample was mounted in a closed-cycle refrigerator with the $\bm{a}$ axis vertical so that the satellites lie along the vertical \textsuperscript{3}He detector banks. For the data in Fig.~1c and Fig.~4 of the main text the sample was mounted in a 10T cryomagnet with the $\bm{b}$ axis vertical, along the field direction. In this geometry only the $(+\delta,0,1)$ satellite was accessible. In both cases, the sample was rotated such that the peak being measured (satellite or commensurate) was located close to forward scattering where the flux at the relevant $d$-spacing (around \SI{19.5}{\angstrom}) is highest. Data analysis was performed with the \textsc{Mantid} software package \cite{Arnold2014}. All data is normalised to the cumulative current and a beam monitor. The intensities were obtained by diffraction focussing a small area on the detector around the peak, and then integrating the resulting time-of-flight spectra.

\section{Resonant elastic x-ray scattering}

\begin{figure*}
	\includegraphics[width=\linewidth]{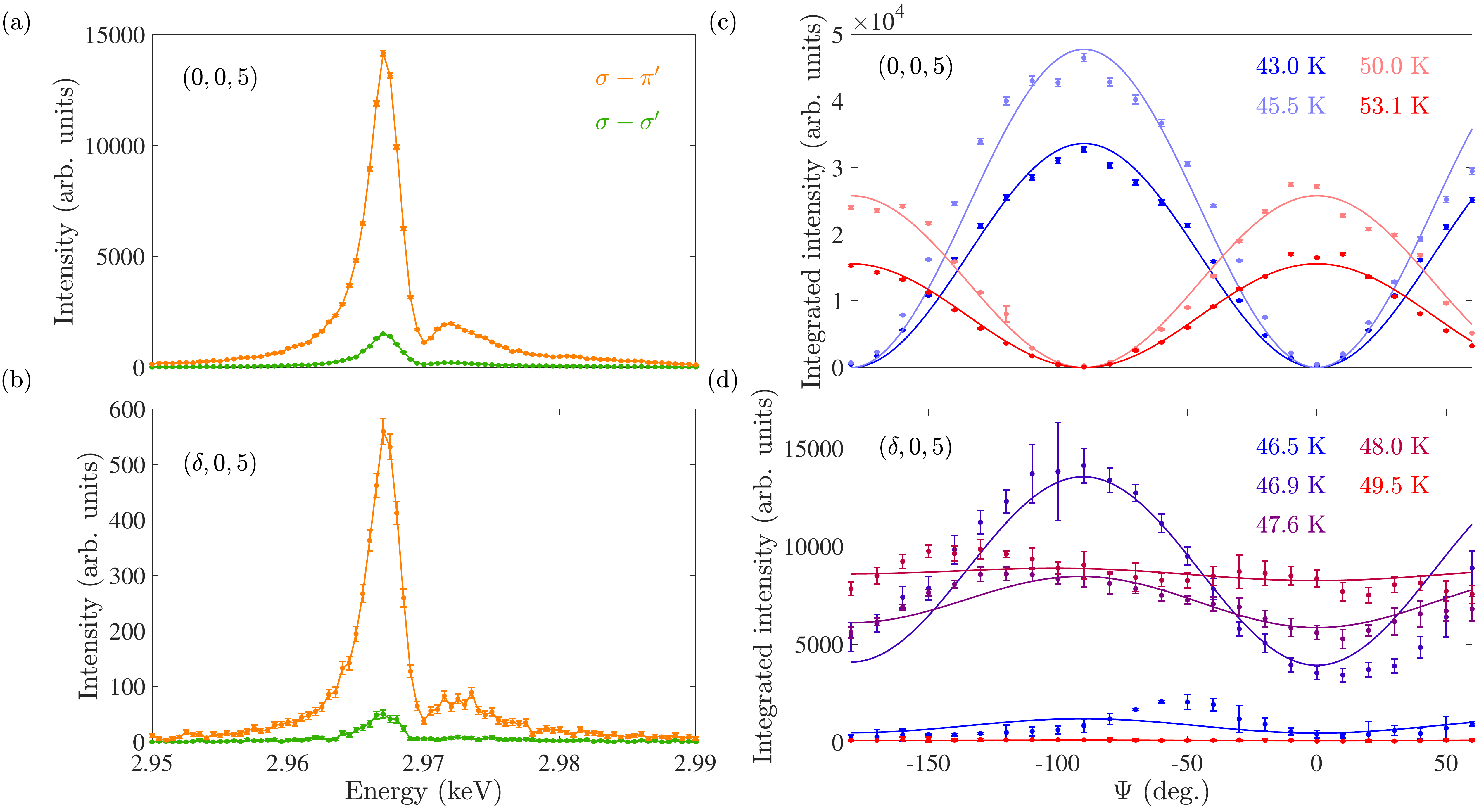}
	\caption{\label{Figure_S2}Additional x-ray data. (a) Energy scans of the $(0,0,5)$ peak at \SI{10.5}{\kelvin} in the $\sigma-\pi'$ (orange) and $\sigma-\sigma'$ (green) polarisation channels, showing the resonant nature of the peak. The small signal in the $\sigma-\sigma'$ channel is due to leakage through the analyser. (b) Energy scans of the $(\delta,0,5)$ peak at \SI{48}{\kelvin} in the two polarisation channels. (c) Azimuthal scans of the $(0,0,5)$ peak at temperatures not shown in the main text. (d) Azimuthal scans of the $(\delta,0,5)$ peak at temperatures not shown in the main text.}
\end{figure*}

Resonant elastic x-ray scattering measurements were performed at beamline I16 of the Diamond Light Source, where a six-circle kappa diffractometer allows a wide range of reciprocal space to be accessed. The horizontally ($\sigma$) polarised incident beam was tuned to the Ru $L_2$ edge (\SI{2.967}{\kilo\electronvolt}) by a double-bounce on a channel-cut Si $(1,1,1)$ monochromator. Polarisation analysis of the scattered beam, shown in Fig.~2a and 3a of the main text, was achieved by a combination of  in-vacuum graphite $(0,0,2)$ analyser and avalanche photodiode detector. Measurement of the direct beam showed leakage between the polarisation channels of $\sim$10\%. The total scattered intensity, shown in the remainder of Fig.~2 and 3 of the main text, was measured using an in-vacuum Pilatus 100K area detector in ultrahigh gain mode. An extended in-vacuum beam pipe was used to reduce air absorption of the beam. The diffractometer was operated in fixed-azimuth mode with a vertical scattering geometry, as indicated in Fig.~2b of the main text. The sample was mounted to a copper holder with GE varnish to ensure good thermal contact and no thermally-induced strain, and was cooled with an Advanced Research System closed-cycle cryocooler. Data analysis was carried out using the \textsc{Py16} program written by D.G.P.~\cite{Py16}. All data is normalised to the ring current and corrected for self-absorbtion (this has minimal effect on the azimuthal dependences as the beam-footprint is nearly constant so close to the specular condition). Intensities in the azimuthal dependences are obtained from summation over a region-of-interest on the area detector, and then fitting of the resulting rocking curves with pseudo-Voigt profiles plus a constant background. This provides full three-dimensional integration of the peaks with virtually zero background contribution.

\section{Free energy invariants}

In this section, we categorise the symmetry of the high temperature (\AFMa{}) and low temperature (\AFMb{}) collinear commensurate antiferromagnetic structures, and the intermediate incommensurate cycloidal (ICC) antiferromagnetic structure found in \CaRuO{}. We then present a Lifshitz-type free energy invariant in terms of the \AFMa{} and \AFMb{} magnetic order parameters, which describes an incommensurate instability originating in Dzyaloshinskii-Moriya antisymmetric exchange that stabilises the spontaneous cycloidal order. Finally in this section, we give free energy invariants that generate cycloidal harmonics when an in-plane magnetic field is applied to the cycloid, transverse to the direction of propagation. These harmonics combine to form a phase-modulated cycloid described in the next section.

We employ the standard setting of the paramagnetic space group, $Cmc2_1$, throughout. To transform between the non-standard space group setting $Bb2_1m$ used in the literature and the standard setting one can simply rotate the direct lattice basis:
\begin{eqnarray}
	a(Bb2_1m)\rightarrow b(Cmc2_1) \nonumber \\
	b(Bb2_1m)\rightarrow c(Cmc2_1) \nonumber \\
	c(Bb2_1m)\rightarrow a(Cmc2_1) \nonumber 
\end{eqnarray}

Both \AFMa{} and \AFMb{} magnetic structures have Y-point propagation vector $\bm{k} = (1,0,0)$. The full Y-point magnetic representation for the Ru Wyckoff site $8b$ decomposes into four one-dimensional irreducible representations (irreps): mY$_1$, mY$_2$, mY$_3$, and mY$_4$. Symmetry-adapted modes of mY$_4$ correspond to the \AFMa{} magnetic structure, and symmetry-adapted modes of mY$_2$ correspond to \AFMb{}.

Consider two, one-dimensional magnetic order parameters, $\mu$ and $\rho$, that transform by the mY$_2$ and mY$_4$ irreps, respectively. These two order parameters can be combined into a Lifshitz-type free energy term that is symmetry allowed by the non-centrosymmetric $Cmc2_11'$ space group:
\begin{equation}
	\mu \frac{\partial \rho}{\partial y} - \rho \frac{\partial \mu}{\partial y}
\end{equation}
where the component of the gradient operator, $\frac{\partial}{\partial y}$, transforms as a polar vector $\parallel y$ ($\Gamma_3$ irrep). The symmetry invariance of this term can be appreciated by inspection of the matrix form of the $Cmc2_11'$ generators (Tables \ref{Table_1} and \ref{Table_2}), which demonstrate the transformational properties of order parameters associated with the mY$_2$, mY$_4$, and $\Gamma_3$ irreps. It is well known that the presence of Lifshitz type invariants in the free energy implies an instability of the system towards inhomogeneous modulated states \cite{Izyumov1990}. For instance, this instability is responsible for the formation of a long-period cycloidal magnetic structure in polar BiFeO\textsubscript{3} \cite{Sosnowska1995, Kadomtseva2004}. The free energy term specified above describes an incommensurate instability away from the commensurate order of \AFMa{} and \AFMb{}, and promotes a cycloidal order propagating along the $\bm{b}$-axis with propagation vector $\bm{k} = (1,\delta,0)$.

\begin{table*}
	\caption{\label{Table_1}Irreducible representation (irrep) matrices for rotation and mirror generators of the $Cmc2_11'$ space group and $T$, the time reversal operator ($\Gamma$-point, $\bm{k}=(0,0,0)$; Y-point, $\bm{k}=(1,0,0)$; $\Delta$-line of symmetry, $\bm{k}=(0,k_y,0)$).}
	\begin{ruledtabular}
		\begin{tabular}{c | c | c c c c c c}
			& Irrep & $\{1|0,0,0\}$ & $\{2_z|0,0,1/2\}$ & $\{m_x|0,0,0\}$ & $\{m_y|0,0,1/2\}$ & T \\
			\hline
			$\mu$ & mY$_2$ & 1 & 1 & -1 & -1 & -1 \\
			$\rho$ & mY$_4$ & 1 & -1 & -1 & 1 & -1 \\
			$\frac{\partial}{\partial y}$ & $\Gamma_3$ & 1 & -1 & 1 & -1 & 1 \\
			H$_z$ & m$\Gamma_2$ & 1 & 1 & -1 & -1 & -1\\
			$(\eta,\eta^*)$ & m$\Delta_2$ & $\begin{pmatrix} 1 & 0 \\ 0 & 1 \end{pmatrix}$ & $\begin{pmatrix} 0 & 1 \\ 1 & 0 \end{pmatrix}$ & $\begin{pmatrix} -1 & 0 \\ 0 & -1 \end{pmatrix}$ & $\begin{pmatrix} 0 & -1 \\ -1 & 0 \end{pmatrix}$ & $\begin{pmatrix} -1 & 0 \\ 0 & -1 \end{pmatrix}$
		\end{tabular}
	\end{ruledtabular}
\end{table*}

\begin{table*}
	\caption{\label{Table_2}Irreducible representation (irrep) matrices for translation generators of the $Cmc2_11'$ space group.}
	\begin{ruledtabular}
		\begin{tabular}{c | c | c c c c c}
			& Irrep & $\{1|1/2,1/2,0\}$ & $\{1|1,0,0\}$ & $\{1|0,1,0\}$ & $\{1|0,0,1\}$ \\
			\hline
			$\mu$ & mY$_2$ & -1 & 1 & 1 & 1 \\
			$\rho$ & mY$_4$ & -1 & 1 & 1 & 1 \\
			$\frac{\partial}{\partial y}$ & $\Gamma_3$ & 1 & 1 & 1 & 1 \\
			H$_z$ & m$\Gamma_2$ & 1 & 1 & 1 & 1 \\
			$(\eta,\eta^*)$ & m$\Delta_2$ & $\begin{pmatrix} \mathrm{e}^{-\pi i k_y} & 0 \\ 0 & \mathrm{e}^{\pi i k_y} \end{pmatrix}$ & $\begin{pmatrix} 1 & 0 \\ 0 & 1 \end{pmatrix}$ & $\begin{pmatrix} \mathrm{e}^{-2\pi i k_y} & 0 \\ 0 & \mathrm{e}^{2\pi i k_y} \end{pmatrix}$ & $\begin{pmatrix} 1 & 0 \\ 0 & 1 \end{pmatrix}$
		\end{tabular}
	\end{ruledtabular}
\end{table*}

The incommensurate propagation vector $\bm{k} = (1,\delta,0)$ is equivalent to $\bm{k} = (0,\delta-1,0)$ located on the $\Delta$-line of symmetry. The reducible $\Delta$-line magnetic representation for the Ru Wyckoff site decomposes into two two-dimensional irreps, m$\Delta_1$, and m$\Delta_2$. The compatibility relationships for m$\Delta_1$, m$\Delta_2$, and the Y-point irreps,
\begin{equation}
	\mathrm{m}\Delta_1 = \mathrm{m}\mathrm{Y}_1 \oplus \mathrm{m}\mathrm{Y}_3
\end{equation}
\begin{equation}
	\mathrm{m}\Delta_2 = \mathrm{m}\mathrm{Y}_2 \oplus \mathrm{m}\mathrm{Y}_4
\end{equation}
indicate that the modulated states that transform by the m$\Delta_2$ irrep decompose into mY$_2$ and mY$_4$ order parameters in the commensurate limit. Hence, the magnetic cycloid stabilised by the Lifshitz invariant, which is expected to be a small perturbation of the commensurate magnetic structures associated with the mY$_2$ and mY$_4$ irreps, transforms according to the m$\Delta_2$ irrep (listed in Tables \ref{Table_1} and \ref{Table_2}).

In the zero magnetic field scattering experiments the spontaneous cycloidal order was found to have a single (fundamental) propagation vector, $\pm \bm{k}_1 \approx (0,\pm0.97,0)$, where the subscript denotes that this is the fundamental (1\textsuperscript{st} order) propagation vector. We write the general complex order parameter of the fundamental cycloidal modulation as $(\eta_1,\eta_1^*)$, where $\eta_1$ and $\eta_1^*$ are complex and complex conjugate components corresponding to $+\bm{k}_1$ and $-\bm{k}_1$, and their appropriate linear combination provides real values for the ruthenium magnetic moments.

We now consider the effect of applying a magnetic field in the plane of the cycloid, and transverse to the direction of propagation ($H_z$). $H_z$, and the $\Gamma$-point magnetisation it creates, transform by the m$\Gamma_2$ irrep, as listed in Tables \ref{Table_1} and \ref{Table_2}. The presence of the field implies a coupling of additional magnetic order parameters, with the lowest order coupling invariant that is linear in $H_z$ being
\begin{equation}
	H_z (\eta_2^* \eta_1^2 + \eta_2 \eta_1^{*2})
\end{equation}
where $(\eta_2,\eta_2^*)$ is a second harmonic cycloidal component with propagation vector $\bm{k}_2 = 2\bm{k}_1$, which also transforms as m$\Delta_2$. This invariant dictates that, when $H_z$ is finite, the energy of the system is lowered through the creation of a second harmonic modulation. More generally, this invariant can be written to higher orders as
\begin{equation}
	H_z (\eta_n^* \eta_1^n + \eta_n \eta_1^{*n})
\label{Equation_even_invariant}
\end{equation}
where $(\eta_n,\eta_n^*)$ is the two-dimensional order parameter of the $n^\mathrm{th}$ harmonic cycloid, and $n$ = 2, 4, 6, ... \textit{i.e.} an infinite series of even harmonics with propagation vector $\bm{k}_n = n \bm{k}_1$ are created by this free energy invariant.

Similarly, odd order harmonics can be expected to occur in the field-modulated cycloid, and they are accounted for by invariants that are quadratic rather than linear in the field:
\begin{equation}
	H_z^2 (\eta_n^* \eta_1^n + \eta_n \eta_1^{*n})
\label{Equation_odd_invariant}
\end{equation}
where $n$ = 3, 5, 7, ...

These field-induced harmonics combine to form a phase modulated cycloid, described in the following section.

\section{Phase modulated cycloid}

In this section, we provide an analytical model of a phase modulated cycloid and its Fourier decomposition, valid for small phase modulations. We consider a cycloid with moments rotating in the $\bm{x}-\bm{y}$ plane, propagating along $\bm{x}$, and with an external magnetic field applied along $\bm{y}$. For simplicity, we take a single atom in the unit cell, which is sufficient to demonstrate the general Fourier decomposition of the cycloid, but not sufficient for structure factor calculations of the neutron and x-ray diffraction intensity from the real material which has eight Ru atoms in the conventional unit cell.

\begin{figure*}
	\includegraphics[width=\linewidth]{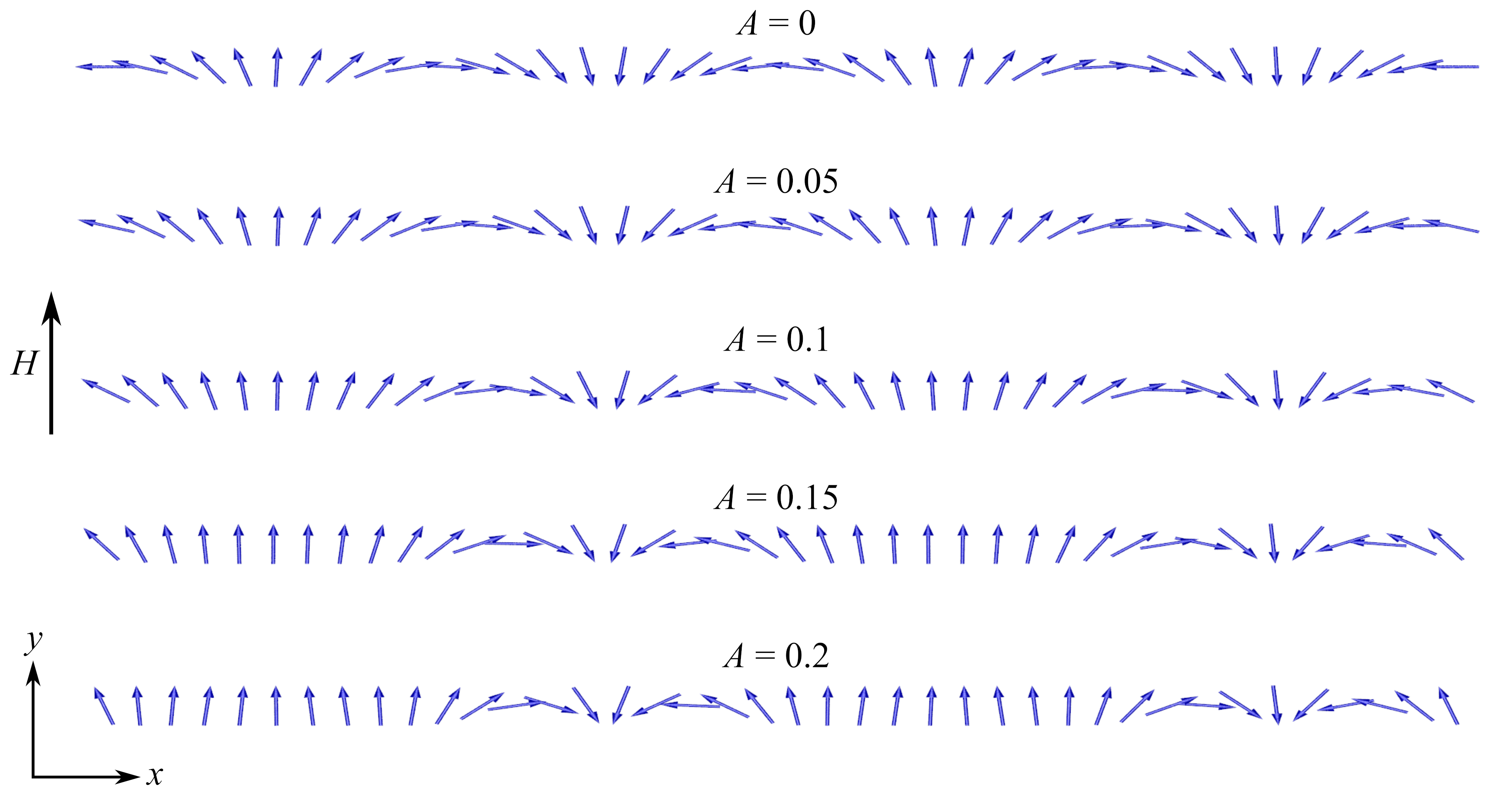}
	\caption{\label{Figure_S3}$\mathcal{M}$ drawn with a circular envelope for different values of the phase modulation amplitude $A$.}
\end{figure*}

We can write the magnetic moment, $\mathcal{M}$, of a given atom as
\begin{equation}
	\mathcal{M}^{\bm{r}} = m_x^{\bm{r}} \hat{\bm{x}} + m_y^{\bm{r}} \hat{\bm{y}}
	\label{Equation_M}
\end{equation}
where $\bm{r}$ is a direct lattice vector that relates the atom's unit cell to the origin. For a phase modulated cycloid with phase modulation periodicity the same as that of the fundamental cycloid (as expected for a $\Gamma$-point field induced modulation), we define
\begin{equation}
	m_x^{\bm{r}} = M_x \cos{(\phi_{\bm{r}} + 2 \pi A \cos{(\phi_{\bm{r}})})}
	\label{Equation_mx}
\end{equation}
\begin{equation}
	m_y^{\bm{r}} = M_y \sin{(\phi_{\bm{r}} + 2 \pi A \cos{(\phi_{\bm{r}})})}
	\label{Equation_my}
\end{equation}
where $\phi_{\bm{r}}= 2 \pi \bm{k}_1 \cdot \bm{r}$, $M_x$ and $M_y$ are the moment magnitudes along $\bm{x}$ and $\bm{y}$ that define an elliptical envelope for the rotating moments, and $A$ is the amplitude of the phase modulation. Figure \ref{Figure_S3} illustrates $\mathcal{M}$ with a circular envelope for different values of $A$. One can observe the magnetisation of the cycloid parallel to the applied field that is established via the phase modulation (changing the sign of $A$ corresponds to the opposite field direction).

By writing Equations \ref{Equation_mx} and \ref{Equation_my} as
\begin{equation}
	m_x^{\bm{r}} = M_x [\cos{(\phi_{\bm{r}})} \cos{(2 \pi A \cos{(\phi_{\bm{r}})})} - \sin{(\phi_{\bm{r}})} \sin{(2 \pi A \cos{(\phi_{\bm{r}})})}]
\end{equation}
\begin{equation}
	m_y^{\bm{r}} = M_y [\sin{(\phi_{\bm{r}})} \cos{(2 \pi A \cos{(\phi_{\bm{r}})})} + \cos{(\phi_{\bm{r}})} \sin{(2 \pi A \cos{(\phi_{\bm{r}})})}]
\end{equation}
and expanding the factors dependent on the phase modulation amplitude $A$ as a Taylor series (in the following we expand cos($x$) up to $x^4$ and sin($x$) up to $x^3$), one can show that
\begin{equation}
	m_x^{\bm{r}} = M_x [\alpha_1 \cos{(\phi_{\bm{r}})} + \alpha_2 \sin{(2 \phi_{\bm{r}})} + \alpha_3 \cos{(3 \phi_{\bm{r}})} + \alpha_4 \sin{(4\phi_{\bm{r}})} + ...]
\end{equation}
\begin{equation}
	m_y^{\bm{r}} = M_y [\beta_0 + \beta_1 \sin{(\phi_{\bm{r}})} + \beta_2 \cos{(2 \phi_{\bm{r}})} + \beta_3 \sin{(3 \phi_{\bm{r}})} + \beta_4 \cos{(4 \phi_{\bm{r}})} + ...]
\end{equation}
where
\begin{eqnarray}
	\alpha_1 &=& 1 - \frac{3}{2} (\pi A)^2 + \frac{5}{12} (\pi A)^4 \\ 
	\alpha_2 &=& \frac{1}{3} (\pi A)^3 - (\pi A) \\
	\alpha_3 &=& \frac{5}{24} (\pi A)^4 - \frac{1}{2} (\pi A)^2 \\
	\alpha_4 &=& \frac{1}{6} (\pi A)^3
\end{eqnarray}
and
\begin{eqnarray}
	\beta_0 &=& (\pi A) - \frac{1}{2} (\pi A)^3 \\
	\beta_1 &=& 1 - \frac{1}{2} (\pi A)^2 + \frac{1}{12} (\pi A)^4 \\
	\beta_2 &=& (\pi A) - \frac{2}{3} (\pi A)^3 \\
	\beta_3 &=& \frac{1}{8} (\pi A)^4 - \frac{1}{2} (\pi A)^2 \\
	\beta_4 &=& -\frac{1}{6} (\pi A)^3 
\end{eqnarray}

If $A = 0$, only the $\alpha_1$ and $\beta_1$ coefficients of the fundamental modulation are non-zero, as expected. $M_y \beta_0$ is exactly the magnetisation of the cycloid parallel to the field direction, $\bm{y}$, which, to first order, is directly proportional to the phase modulation amplitude $A$. Furthermore, we note that the coefficients of the odd harmonics are in terms of only even powers of $A$, and the coefficients of the even harmonics are in terms of only odd powers of $A$. This is fully consistent with the free energy invariants given in Equations \ref{Equation_even_invariant} and \ref{Equation_odd_invariant}, by which the even and odd harmonics couple linearly and quadratically to the magnetic field, respectively.

\begin{table}
	\caption{\label{Table_3}Fourier coefficients $\psi_n$ and $\chi_n$ and their relationship to $\alpha_n$ and $\beta_n$. The amplitudes and intensities have been evaluated for $A = 0.1$.}
	\begin{ruledtabular}
		\begin{tabular}{c | c c}
			Fourier coefficient & Amplitude & Intensity \\
			\hline
			$\psi_0 = \alpha_0$ & 0 & 0 \\
			$\psi_1 = \alpha_1$ & 0.856 & 0.733 \\
			$\psi_2 = -i \alpha_2$ & 0.304 $i$ & 0.092 \\
			$\psi_3 = \alpha_3$ & -0.047 & 0.002 \\
			$\psi_4 = -i \alpha_4$ & -0.005 $i$ & $\sim 0$\\ 
			\hline
			$\chi_0 = \beta_0$ & 0.299 & 0.089 \\
			$\chi_1 = -i \beta_1$ & -0.951 $i$ & 0.905 \\
			$\chi_2 = \beta_2$ & 0.293 & 0.086 \\
			$\chi_3 = -i \beta_3$ & 0.048 $i$ & 0.002 \\
			$\chi_4 = \beta_4$ & -0.005 & $\sim 0$
		\end{tabular}
	\end{ruledtabular}
\end{table}

Finally, we consider the Fourier transform of the phase modulated cycloid. Equation \ref{Equation_M} can be recast as the sum over Fourier components, $\bm{S}_{\bm{k}}$.
\begin{equation}
	\mathcal{M}^{\bm{r}} = \sum_{\bm{k}} \bm{S}_{\bm{k}} \mathrm{e}^{2 \pi i \bm{k} \cdot \bm{r}}
	\label{Equation_Mf}
\end{equation}
where
\begin{equation}
	\bm{S}_{\bm{k}} = \frac{1}{2} (\psi_n M_x \hat{\bm{x}} + \chi_n M_y \hat{\bm{y}})
\end{equation}
The sum in Equation \ref{Equation_Mf} must be taken over $\pm \bm{k}$, and $\bm{S}_{\bm{k}} = \bm{S}_{\bm{k}}^*$.
The Fourier coefficients $\psi_n$ and $\chi_n$ are related to $\alpha_n$ and $\beta_n$, as shown in Table \ref{Table_3}. We also tabulate numerical values for the Fourier coefficients (evaluated for $A = 0.1$) and their modulus squared, which gives an indication of the expected, relative diffraction intensities.

\bibliography{Ca327_PRB_SM}